\documentclass[aps,pra,superscriptaddress,twocolumn,showpacs,amsmath,amsfonts,amssymb,floatfix,showkeys]{revtex4-1}
\usepackage{graphicx}
\usepackage{epsfig}
\usepackage{subfigure}
\usepackage{sidecap}
\usepackage{amsmath,amssymb,amsthm}
\usepackage{epstopdf}

\newcommand{\Hc }{\mathcal{H}}      
\usepackage{hyperref}
\begin{document}

\title{Collectively induced many-vortices topology \\ via rotatory Dicke quantum phase transition}

\author{Priyam Das}
   \affiliation{Institute of Nuclear Science, Hacettepe University, Ankara - 06800, Turkey}

\author{Mehmet Emre Tasgin}\thanks{Corresponding author: metasgin@hacettepe.edu.tr}
   \affiliation{Institute of Nuclear Science, Hacettepe University, Ankara - 06800, Turkey}

\author{\"{O}zg\"{u}r E. M\"{u}stecapl\i{o}\u{g}lu}
   \affiliation{Department  of  Physics, Ko\c{c} University,  Sar{\i}yer, \.{I}stanbul, 34450, Turkey}

\begin{abstract}
We examine the superradiance of a Bose-Einstein condensate pumped with a Laguerre-Gaussian laser of high winding number, e.g., $\ell = 7$.  The laser beam transfers its orbital angular momentum (OAM) to the condensate at once due to the collectivity of the superradiance. An $\ell$-fold rotational symmetric structure emerges with the take place of rotatory superradiance. $\ell$ number of single-charge vortices appear at the arms of this structure. Even though the pump and the condensate profiles initially have cylindrical symmetry, we observe that it is broken to $\ell$-fold rotational symmetry during the superradiance. Breaking of the cylindrical symmetry into the $\ell$-fold symmetry and OAM transfer to the BEC become possible after the same critical pump strength. Reorganization of the condensate resembles the ordering in the experiment by Esslinger and colleagues [Nature, {\bf 264}, 1301 (2010)]. We show that the critical point for the onset of the reorganization follows the form of the Dicke quantum phase transition.
\end{abstract}



\maketitle

\section{Introduction} 

Superradiance (SR) is the collective emission of an ensemble of atoms into an initially unoccupied mode \cite{dicke1954,skribanowitz1973observation}. Atoms interact with the common electromagnetic field within the extent of the pump's coherence length, and are excited to many-body states. Vacuum fluctuations in an empty photon mode stimulate the decay (emission) of atoms to this mode \cite{andreev1993cooperative}. Hence, SR is often called as the spontaneous collective emission of an ensemble. This makes SR strongly directional~\cite{Moore1999}. Probability of the ensemble to spontaneously decay to an empty mode is proportional to the exponential of the ensemble length along this direction \cite{Moore1999}.

SR is a quantum phase transition (PT) \cite{emary2003,nagy2010,esslinger2010}, which can be treated with the Dicke model \cite{dicke1954}. Above a critical pump (or effective atom-photon coupling) strength, phase transition results in the macroscopic occupation of the electronic excited states. The collectivity, which  SR induces, leads to bipartite entanglement of particles \cite{brandes2004}, photon-atom entanglement \cite{brandes2004}, single-mode nonclassicality \cite{tasgin2015single} but more importantly, to $N$-particle (collective) entanglement \cite{yukalov2004} of particles via spin-squeezing \cite{sorensen2001}. Spin-squeezing is the collective entanglement of all ($N$) of the ensemble atoms, where the measurement of a single atom influences all the remaining ones instantaneously. In a Bose-Einstein condensate (BEC), such a collective entanglement is shown to be induced purely by the atom-atom collisions \cite{sorensen2001}. For the superradiant scattering from a BEC, this preexisting collectivity enables an earlier induction of the phase transition since BEC is already in the symmetric superradiant Dicke states \cite{dicke1954}.

When a BEC is stirred, it does not rotate until a total of $N\hbar$ orbital angular momentum (OAM) is transferred to it \cite{pethick2002}. Experiments on the transfer of OAM to BECs via optical pumping \cite{tabosa1999,anderson2006,wright2008,moretti2009}  make us understand that there exist a critical threshold for breaking the collective behavior of $N$-particle entanglement of a BEC \cite{tasgin2011}. A BEC can be made partially rotate, if the recoil energy  ($\hbar \omega_{R}$), a single atom gains, exceeds the mean atom-atom interaction energy ($U_{int}/N$) which is the situation for the optical frequencies. Therefore, regular optical pumping cannot transfer BEC to a rotatory state collectively.

\begin{figure}
\begin{center}
\includegraphics[width=7.5cm]{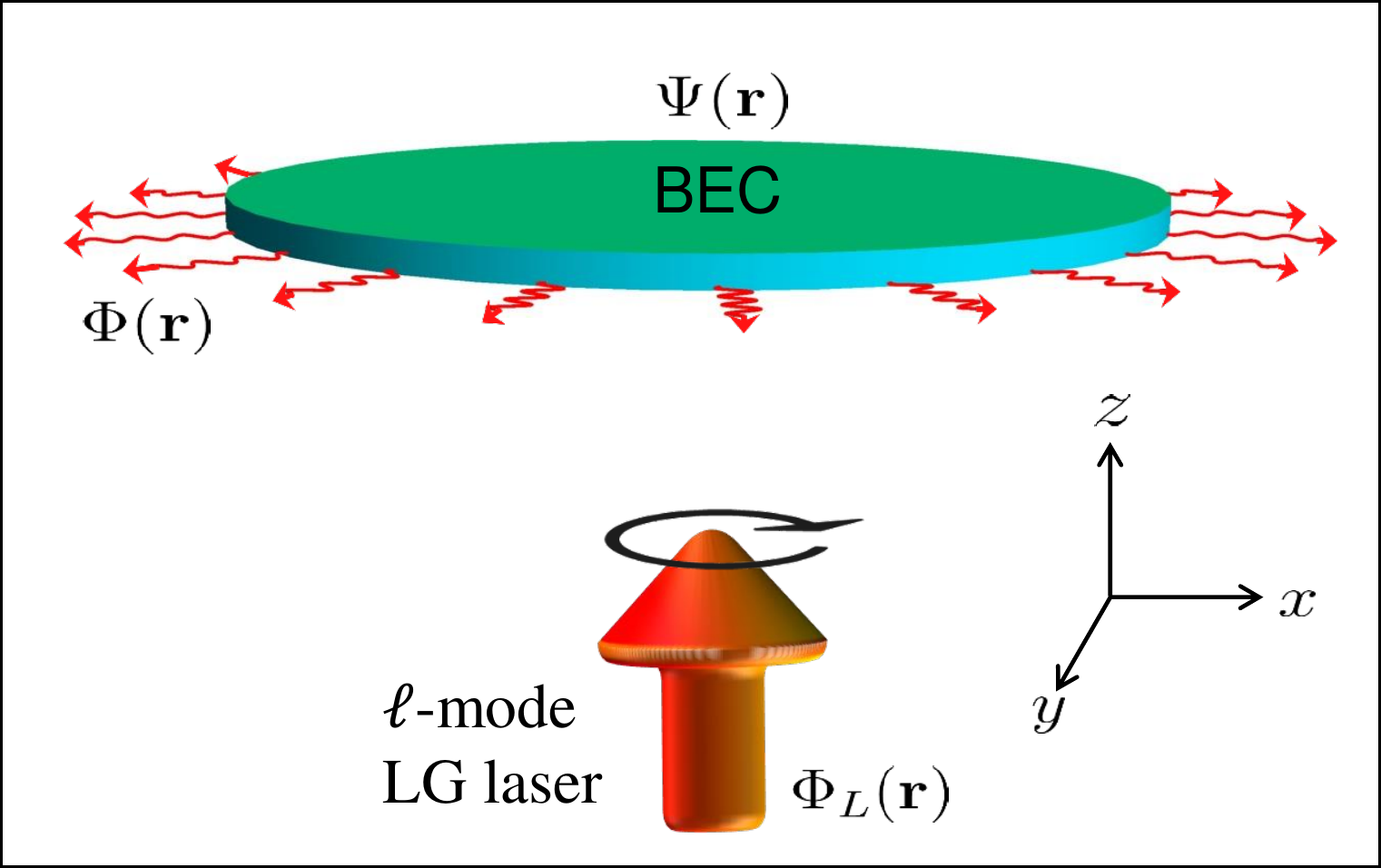}  
\caption{A pancake shaped BEC is illuminated with a strong LG laser of winding number $\ell$. Profile of the laser, Eq.~(\ref{PhiL}) carries $\ell\hbar$ amount of OAM per photon. Above a critical pump strength $\eta>\eta_c$ BEC superradiates in the x-y plane due to the confinement along the z-axis~\cite{GoldbartNaturePhys2009}.
}
\label{fig1}
\end{center}
\end{figure}

We use the collectivity \cite{yukalov2004} of rotatory SR \cite{tasgin2011} to optically induce many vortices in a BEC, at a single time. We pump the BEC with a Lauguerre-Gaussian (LG) beam of higher modes (upto $\ell = 7$), see Fig.~\ref{fig1}. We observe that the $\ell$-fold rotational symmetric density structures emerge, Fig.~\ref{fig5}, even though the pump and BEC initially have cylindrical symmetric intensities.  Interestingly, single-charge vortices emerge at the arms of the $\ell$-fold symmetric structures, see Fig.~\ref{fig5}.  We determine the spatial form of the superradiant scattered pulse, $F(x,y)$, evolving with the equations of motion (EOM). Form of the scattered SR field is free of constraints, e.g., no mode expansion is used.

Reorganization of the condensate into $\ell$-fold rotational symmetric structures, resembles the take place of spatial ordering in normal SR, explored by Esslinger and colleagues \cite{esslinger2010}. In Refs.~\cite{esslinger2010,nagy2008self,nagy2010,oztop2013collective}, the appearance of spatial ordering arises due to the collective transfer of linear momentum via normal SR. Here, $\ell$-fold rotational ordering (and quantized vorticity) takes place due to the excess OAM transfer to the BEC via rotatory SR. Breaking of the cylindrical symmetry into the $\ell$-fold symmetry and OAM transfer to the BEC become possible after the same critical pump strength. Quantized rotations (single-charge vortices) also emerge in an $\ell$-fold rotational ordered form. We observe that rotatory SR emerges without a preceding normal SR, unlike Ref.~\cite{tasgin2011}, when the damping is small.

Onset (dynamics) of a superradiant spontaneous decay can be demonstrated only within the second-quantized scheme \citep{Moore1999,mustecapliouglu2000superradiant}, since it is triggered by vacuum fluctuations. Nevertheless PT manifests itself also in semiclassical treatments as instabilities \cite{nagy2008self} on which experimental results \cite{esslinger2010} can be mapped consistently. 

We show that (i) angular momentum transferred to the BEC, $\langle\hat{L}_z\rangle$, becomes nonzero only after a critical pump strength $\eta>\eta_c$, see Fig.~\ref{fig6}, which is accompanied by the breaking of the cylindrical symmetry into the $\ell$-fold rotational symmetry, see Fig.~\ref{fig7}. (ii) Temporal width of the scattered pulse peak is inversely proportional to the number of atoms~\cite{mandel&wolf_book,bonifacio1975cooperative}, see Fig.~\ref{fig3}, which is characteristic to SR. For $\eta$ sufficiently larger than $\eta_c$, scattered field intensity displays in-phase scattering, $I\sim N^2$, behavior. (iii) Transition is in the Dicke form such that critical pump strength ($\eta_c$) depends on the frequency of the end-fire mode ($\Delta$) and diagonal atom-field coupling ($U_0$) in the form suggested by Nagy {\it et al}. \cite{nagy2008self,esslinger2010}, see Fig.~\ref{fig8}.

\section{Hamiltonian}

We examine the dynamics of a pancake BEC~\cite{GoldbartNaturePhys2009}, see Fig.~\ref{fig1}, which is illuminated with a strong LG laser of optical frequency $\omega_L$ along the symmetry axis $z$. LG pump has a spatial profile
\begin{eqnarray}
\Phi_{L}(\mathbf{r}) = E_L \:  \left(r/w_L\right)^\ell e^{-r^2/2w_L^2}e^{i \ell \phi}e^{ik_L z}
\label{PhiL}
\end{eqnarray}
and carries an $\ell\hbar$ amount of OAM per photon. Detuning enables the adiabatic elimination of the electronic excited state~\cite{meystrePRA1999,tasginthesis}. $w_L$ is the radial width and $k_L$ is the wave-vector of the LG pump.  Electric field amplitude of the laser is $E_L=\alpha_L\varepsilon_L$ where $|\alpha_L|^2$  is the number of pump photons and $\varepsilon_L=(\hbar\omega_L/\epsilon_0 V)^{1/2}$ with $V$ the volume of the laser cavity.

Similar to the end-fire modes emitted out of a pencil shaped condensate~\cite{Moore1999}, scattered radiation off a pancake BEC would consist of radial edge-fire modes, in the xy-plane shown in Fig.~\ref{fig1}.
End-fire modes are emitted through the edge, along the ${\bf s}=x\hat{x}+y\hat{y}$ directions since the BEC is tightly confined along the z-axis. When the x-y profile of the scattered radiation has a smaller winding number compared to the pump winding, OAM is transferred to the BEC. If the scattering is collective, that is superradiant, OAM is transferred to the BEC collectively at a single time, i.e. via $e^{Nt/\tau}$ emission behavior \cite{Moore1999}.

The second quantized effective Hamiltonian, $\hat{\Hc}$, of the system can be written as the sum
\begin{equation}
\hat{\Hc}=\hat{\Hc}_{\rm 0} + \hat{\Hc}_{\rm col} + \hat{\Hc}_{f} + \hat{\Hc}_{af} + \hat{\Hc}_{af}^{(2)}
\label{Htot}
\end{equation}
of the energy of the external (motional) states of the BEC
\begin{equation}
\hat{\Hc}_{0}=\int d^{3}{\bf r} \hat{\Psi}^{\dagger}({\bf r}) \hat{H}_{g} \hat{\Psi}({\bf r}) \: ,
\end{equation}
interatomic collision Hamiltonian
\begin{equation}
\hat{\Hc}_{\rm col}=
g_{s} \int d^{3}{\bf r} \hat{\Psi}^{\dagger}({\bf r}) \hat{\Psi}^{\dagger}({\bf r}) \hat{\Psi}({\bf r}) \hat{\Psi}({\bf r}) \: ,
\label{H0}
\end{equation}
the total electromagnetic energy of  the scattered field
\begin{equation}
\hat{\Hc}_{f} =  \hbar \omega_{e} \int d^{3}{\bf r} \hat{\Phi}^{\dagger}({\bf r}) \hat{\Phi}({\bf r}) \: , 
\label{Hf}
\end{equation}
coupling of the atoms with the pump (off-diagonal term)
\begin{eqnarray}
\hat{\Hc}_{af} = \hbar g_{a} \int d^{3}{\bf r} \hat{\Psi}^{\dagger}({\bf r}) \hat{\Phi}^{\dagger}(r)  \Phi_{L}({\bf r}) \hat{\Psi}({\bf r}) + H.c. \: ,
\label{Haf}
\end{eqnarray}
and the interaction of the scattered field with the atoms (diagonal term)
\begin{eqnarray}
\hat{\Hc}^{(2)}_{af} = 2 \hbar g_{a} \int d^{3}{\bf r} \hat{\Psi}^{\dagger}(r) \hat{\Phi}^{\dagger}(r) \hat{\Phi}({\bf r}) \hat{\Psi}({\bf r}),
\label{Haf2}
\end{eqnarray}
where we make parametric pump approximation for the LG laser. Here, $\hat{H}_g=-\hbar^2\nabla^2/2m+V(r)$ is the first-quantized Hamiltonian for the BEC with $V(r)$ is the trap potential. $\hat{\Phi}(\bf r)$ is the field operator for the superradiantly scattered pulse, whose spatial behavior is determined by the EOM. $\hat{\Psi}({\bf r})$ is the field operator for the BEC. $g_a$ is the effective strength for the atom-field interaction~\cite{meystrePRA1999,tasginthesis}.

Due to the symmetry of the system and the form of the pump~(\ref{PhiL}), we make separation of variables in the BEC and the field operators, $\hat{\Psi}(r)=\hat{f}(x,y)\hat{Z}(z)$ and $\hat{\Phi}(r)=\hat{F}(x,y)Z_f(z)$. This reduces the computational efforts. We obtain the time evolution for each operator using the Heisenberg EOM, e.g., $i\hbar\dot{\hat{F}}(x,y)=[\hat{F},\Hc]$ (see the Appendix). Since we are interested in the field amplitudes, we replace all of the operators by c-numbers, e.g. $\hat{f}(x,y)\to f(x,y)$. 

We scale the BEC wave-function $\Psi({\bf r})$ and the scattered field $\Phi({\bf r})$ with $\sqrt{N}$ \cite{nagy2008self}. Parameters $\eta=g_a\sqrt{N}|\alpha_L|$ and $U_0=g_a N$ control the strength of the off-diagonal ($\hat{\Hc}_{af}$) and diagonal ($\hat{\Hc}_{af}^{(2)}$) atom-field couplings, respectively. Off-diagonal coupling favors the macroscopic occupation the scattered field \cite{emary2003} while energy of the diagonal coupling is minimized with a microscopic scattered field. Hence, the two works against each other. If $g_s$ is negative (positive), collision term supports (works against) the SR transition \cite{nagy2008self,nagy2010}.

Tightly confined BEC does not superradiate along the z-direction. SR field profile along the z-direction nonvanish only in the condensate, which has an extent of $\simeq$10nm that is much smaller than the wavelength of the radiation. Therefore, we neglect $Z_f(z)$ profile in $\Phi({\bf r})$. Additionally, again due to the tight confinement of the BEC along the $z$-direction, vortices pointing along the $x$-$y$ directions are energetically unfavorable. The scattered light (end-fire mode) propagates almost in the $x$-$y$ plane due to the strong directionality of the SR~\cite{Moore1999,inouye1999}, thus has no propagation component along the $z$-direction. 

We note that, in obtaining the EOM we do not make any mode expansion in the BEC, $\Psi({\bf r})$, or the scattered SR field, $\hat{F}(x,y)$, profiles. We aim to observe the free (without constraint) evolution of the scattered field and the BEC through the EOM. This is because, during the time evolution, BEC and the scattered field may attain unexpected profiles, which can maximize/minimize the interaction terms. 

We examine the behavior of the system both in the presence and the absence of damping. Following the mean field theory of free space superradiance from an ensemble of atoms, we introduce a phenomenological linear loss term into the field dynamical equation in our Maxwell-Bloch type equations~\cite{bowden1978kappa,bowden1979kappa,oztop2013collective}, in Eq.~(\ref{App_A1}). Presence of damping causes the 3-fold ordering and the OAM transfer to take place at higher pump intensities, as should be expected.

\section{Vortices with $\ell$-fold rotational symmetry}
We time evolve $F(x,y)$ and $\Psi({\bf r})=f(x,y)Z(z)$ using the EOMs given in the Appendix. In Fig.~\ref{fig2}, macroscopic OAM transfer to the BEC follows the emission peak (at $t\simeq 1.45/\omega_R$). In Fig.~\ref{fig2}(a), the $\langle \hat{L}_z\rangle$ can have values different than pump LG photon's OAM due to collective nature of the superradiant scattering. The temporal width of the peak ($\Delta\tau$), in Fig.~\ref{fig2}(b), follows the superradiant form suggested by the Refs.~\cite{mandel&wolf_book,bonifacio1975cooperative}. When we increase the number of atoms by $n$ times, we observe that temporal width of the peak shrinks to $\Delta\tau/n$, see Fig.~\ref{fig3}. Additionally, the peak intensity follows the form $\propto N^2$ form. Hence, the scattering displays the superradiant character.


\begin{figure}
\begin{center}
\includegraphics[width=8.5cm]{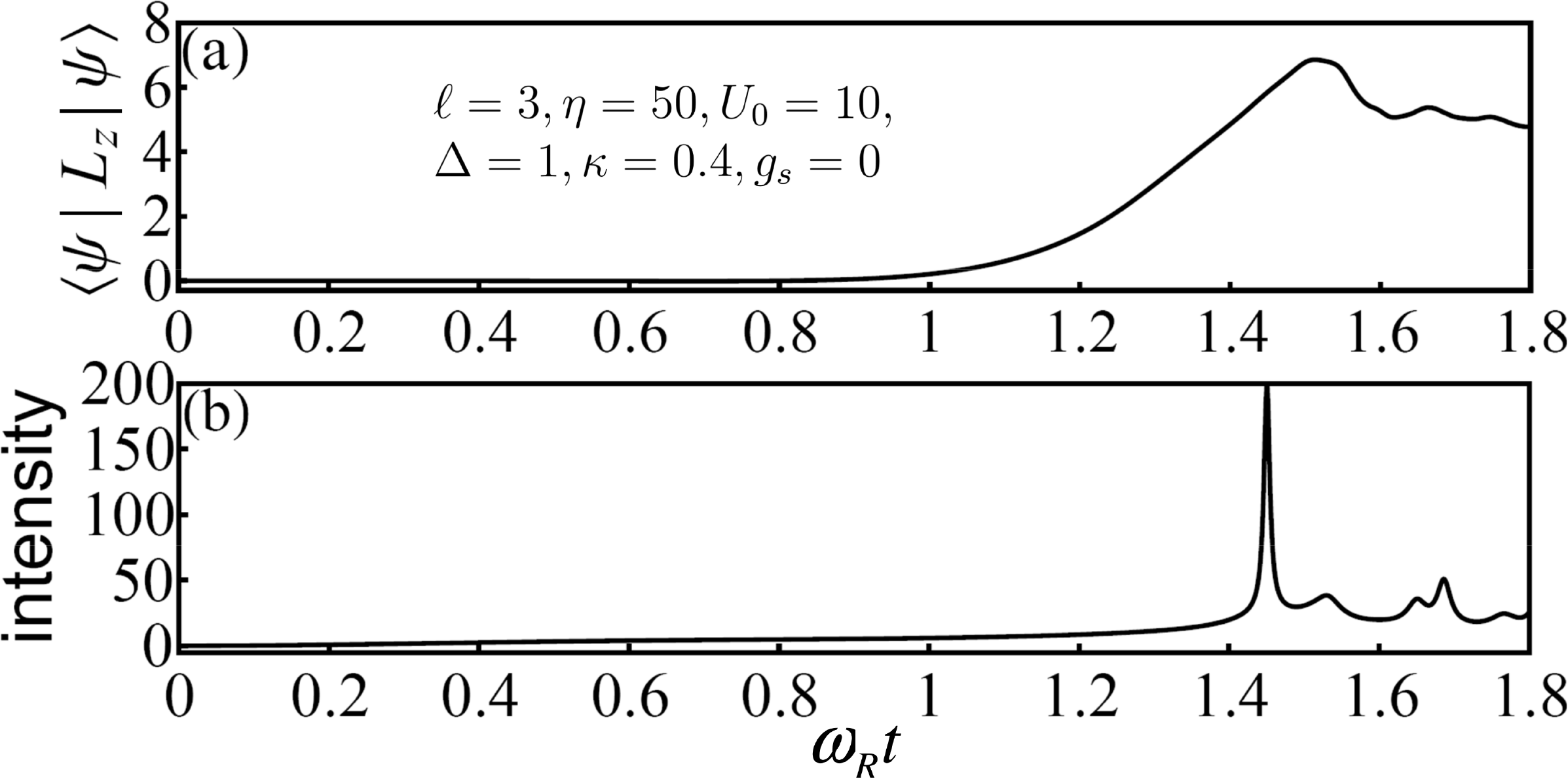}
\caption{(a) OAM transfer to the BEC follows the (b) scattering peak (rotatory SR) at $t\simeq 1.45/\omega_R$. Reorganization of the BEC profile, from azimuthal symmetry to $\ell$-fold rotational symmetry, with $\ell$ quantized vortices, emerges close before the intensity peak.}
\label{fig2}
\end{center}
\end{figure}

\begin{figure}
\begin{center}
\includegraphics[width=8.2cm]{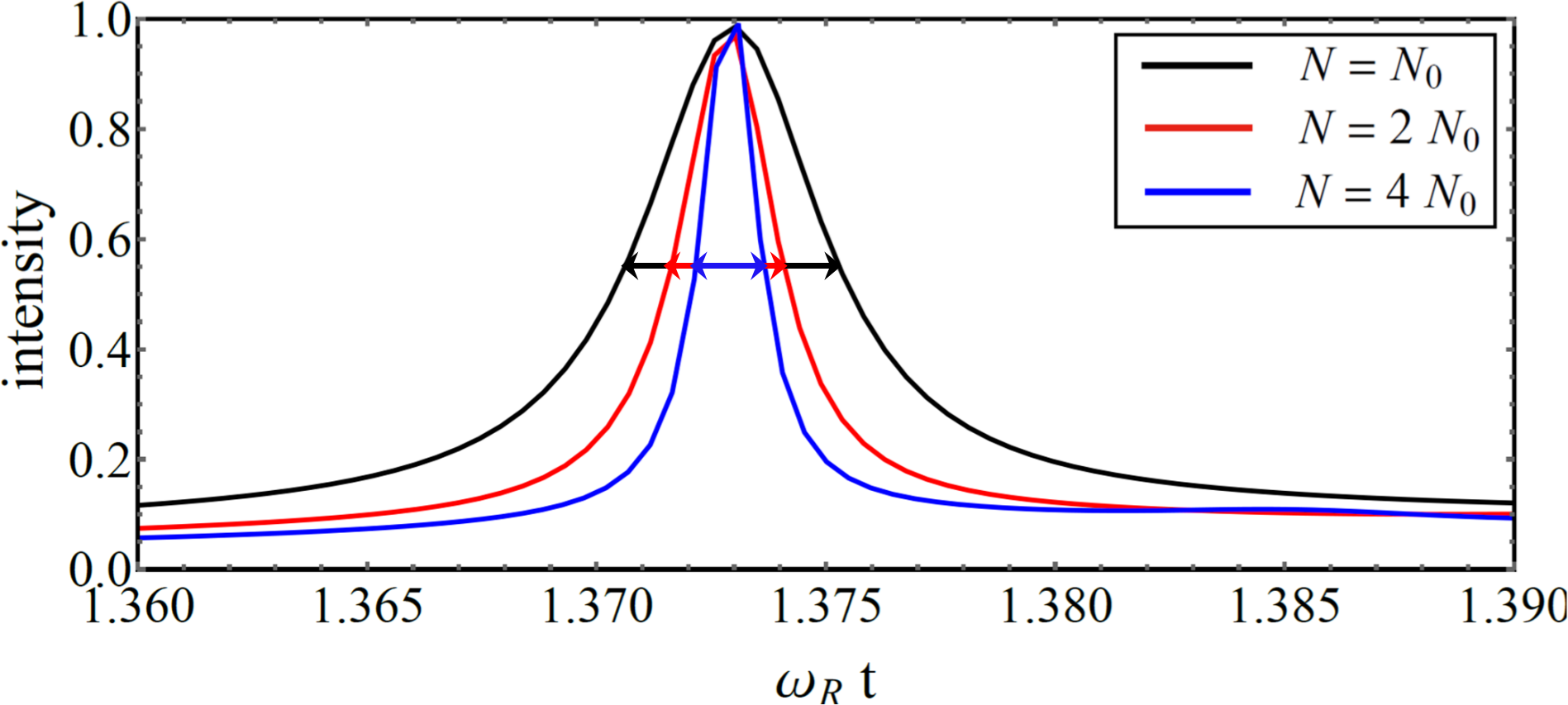}
\caption{The temporal widths of the pulse peaks are proportional to 1/N, as suggested by Mandel \& Wolf~\cite{mandel&wolf_book}. We observe that FWHM is $\Delta\tau$=0.0048 for $N=N_0$; $\Delta\tau$= 0.0023 for $N=2N_0$ and $\Delta\tau$=0.0012 for $N=4N_0$. This is a behavior characteristics to superradiant scattering. We remind that $N$ appears in the parameters as $\eta=g_a\sqrt{N}|\alpha_L|$ and $U_0=g_a N$. (We shifted the three curves to the same peak position and normalized the max intensity to 1 and no damping is assumed.) 
}
\label{fig3}
\end{center}
\end{figure}

\begin{figure}
\begin{center}
\includegraphics[width=	8.0cm]{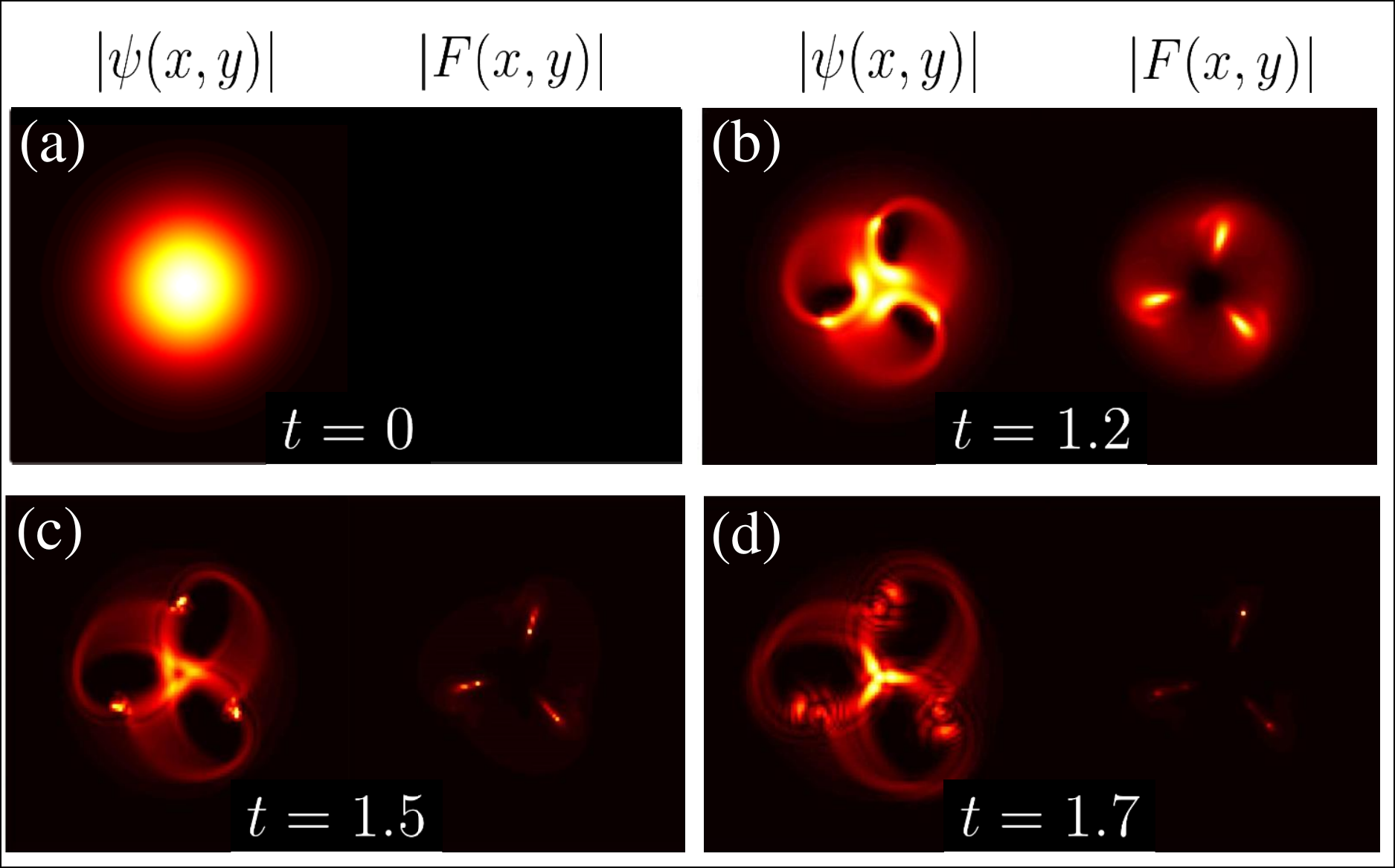}  
\caption{Time evolution of the profiles (corresponding to Fig. 2) of the BEC density $|\psi(x,y)|$ and the scattered field intensity $|F(x,y)|$ for pumping with an $\ell=3$ mode LG laser above the critical pump strength $\eta>\eta_c$. 3-fold rotational symmetric structures appear (b) both in the BEC and scattered field, even though both have cylindrical symmetry initially. (c) Single-charge vortices appear at the ends of the structure such that (d) expansion of the vortices can be observed when the laser is turned off at $t\simeq1.51$. Quantized vortices appear suddenly when $m_{\rm tot}>3$. [In (c) and (d), scattered field are very sharp, so that graphs look as if they are faded due to the scaling of the colormap.]}
\label{fig4}
\end{center}
\end{figure}

In Fig.~\ref{fig4}, we depict the time evolution of the spatial profile for both the BEC and the scattered field. In the close proximity of the scattering peak, Fig.~\ref{fig2}(b), azimuthal symmetric BEC profile reorganizes to 3-fold rotational symmetry, Fig.~\ref{fig4}(b). This happens even though both BEC and LG pump intensity have cylindrical symmetry initially. The winding number of the laser is transformed to the 3-fold rotational symmetry. Three single-charge vortices appear at the arms of the 3-fold rotational symmetric density profile. By enclosing small spatial regions around the three vortex positions, we calculate the expectation value of the $\hat{L}_z$ operator and find that they carry a single charge. In Fig.~\ref{fig4}(d) we observe the expansion of the three vortices when the laser field is lifted off. Remaining OAM $(m_{\rm tot}-\ell)\hbar$, per atom, is distributed to the body of the BEC, where $m_{\rm tot}=\langle\Psi({\bf r})|\hat{L}_z|\Psi({\bf r})\rangle$ is the total OAM of the BEC.


\begin{figure}
\begin{center}
\includegraphics[width=	8.0cm]{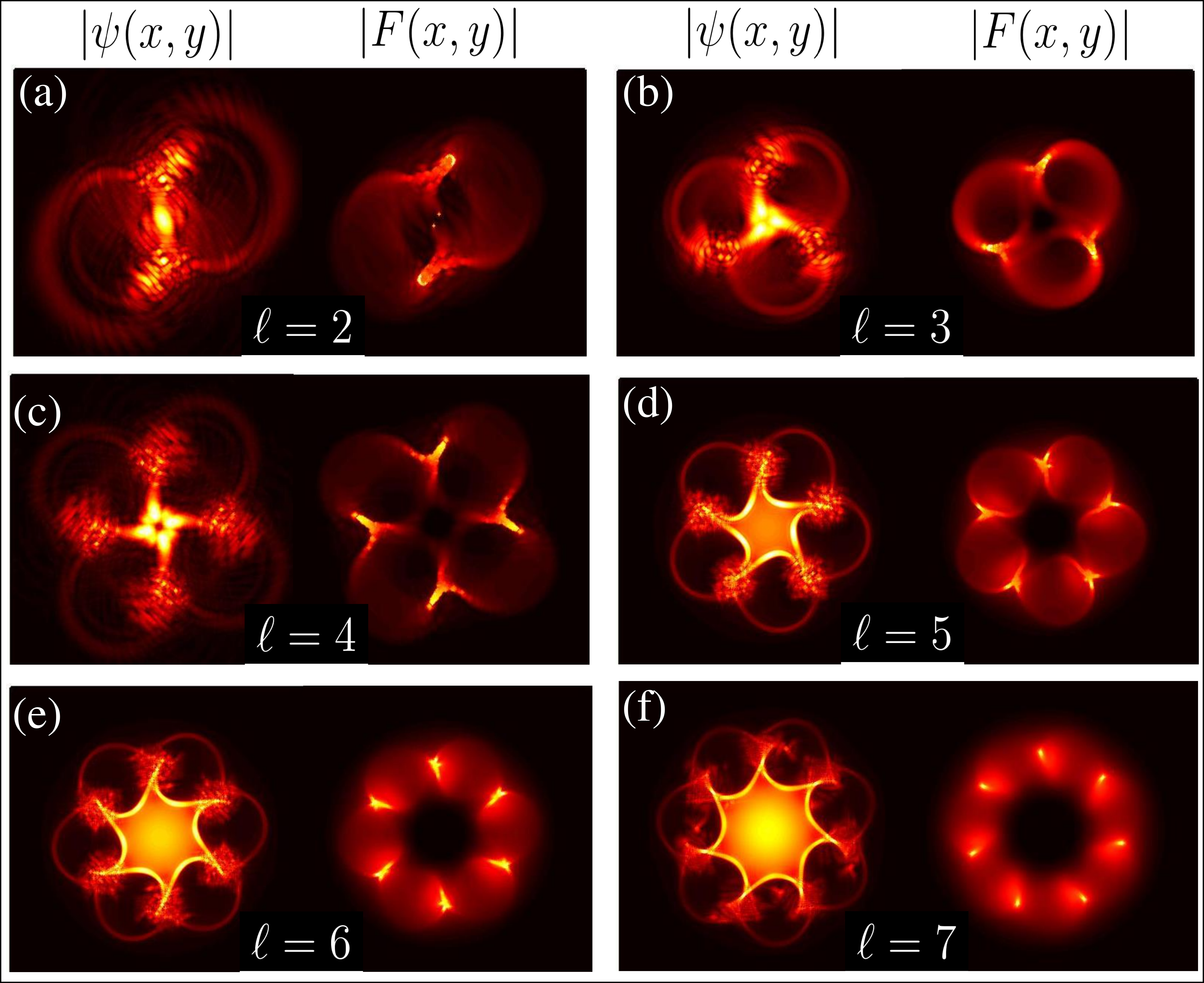} 
\caption{$\ell$-fold rotational symmetric structures emerge for pumping with LG laser, profile (\ref{PhiL}), of winding number $\ell$. $\ell$ number of single-charge vortices appear at the ends of $\ell$-fold structures above a critical pump strength of the $\ell$-mode LG laser. ($\eta$ values are 50, 60, 70, 120, 120, 150 and the times profiles belong to are t=2.2, 1.4, 1.56, 0.39, 0.23, 0.15, respectively, and even a higher value of damping is chosen, $\kappa$=3.)}
\label{fig5}
\end{center}
\end{figure}

In our simulations, $\ell$-fold rotational symmetric structures emerge for the pumping with a LG laser of winding number $\ell$, see Fig.~\ref{fig5}. Single-charge vortices emerge at the ends of $\ell$-fold rotational symmetric structures. When the OAM transferred to BEC is smaller than the $\ell$-fold symmetry, $m_{\rm tot}<\ell$, $\ell$-fold symmetric structures emerge again. However, in this case no single-charge vortex appear at the arms. The whole OAM is distributed to the body of the BEC. No quantized vorticity can be identified in the body of the BEC. 

We note that, in difference to vortex creation via stimulated emission~\cite{BigelowPRA2008,andersen2006OAM}, amounts of OAM which are larger than the one for the pump can be transferred to the BEC via SR. There are two reasons. First, since SR is a collective process, the atoms can absorb and emit multi-photons collectively. Hence, more than one photon per atom can contribute to the collective emission. Second, in our case, the spatial profile of the BEC side-mode in which atoms can recoil is not constrained, since a stimulating laser do not impose the spatial profile of the mode~\cite{BigelowPRA2008,andersen2006OAM}.   

\section{Critical pump strength}

\begin{figure}
\begin{center}
\includegraphics[width=8.5cm]{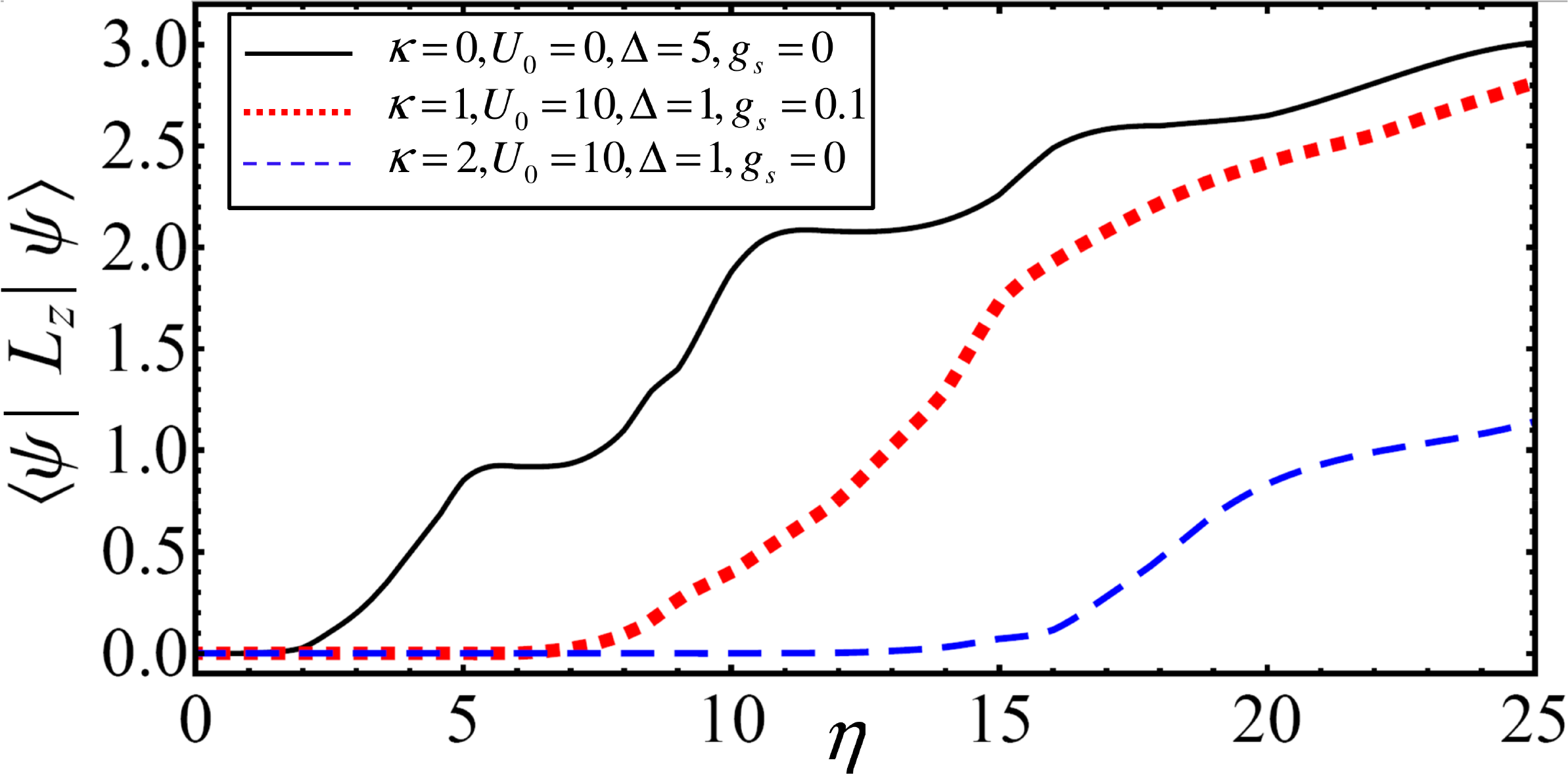}
\caption{OAM transfer to the BEC can be achieved only after a critical pump strength $\eta>\eta_c$, which depends on the actual values of end-fire mode frequency ($\Delta$), diagonal atom-field coupling ($U_0$), damping ($\kappa$) and atom-atom collisions ($g_s$).}
\label{fig6}
\end{center}
\end{figure}

\begin{figure*}
\begin{center}
\includegraphics[width=12cm]{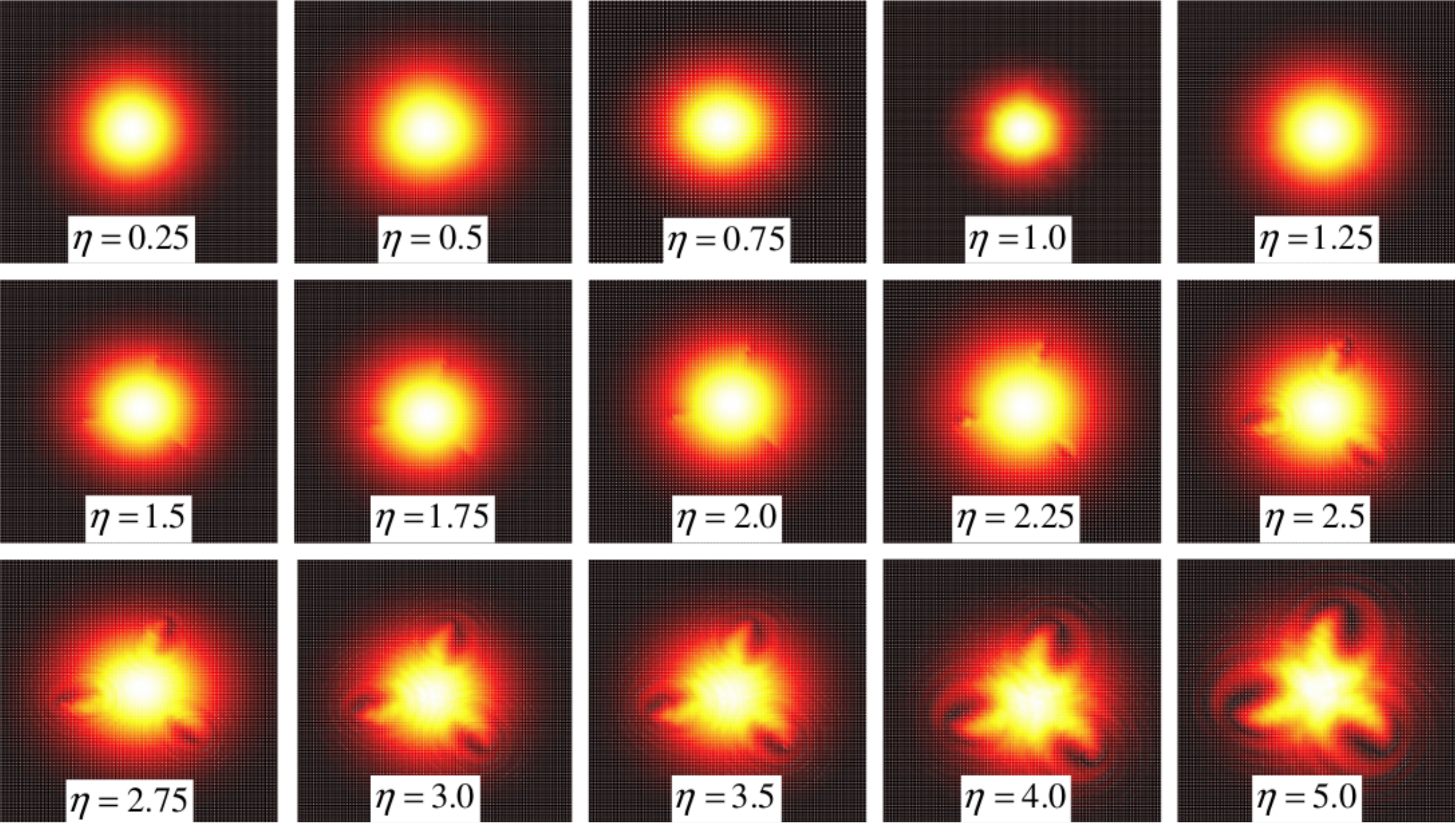}
\caption{Visibility of the 3-fold rotational ordering in the BEC profile ($|F(x,y)|$) for different $\eta$ values. Profiles correspond to solid line in Fig.~\ref{fig6}. For each $\eta$ value, we perform the time evolution and capture the BEC profile where 3-fold ordering is most observable. From $\eta=\eta_c\simeq $2.25  to $\eta$=2.5, the visibility of the 3-fold symmetry jumps. Only after $\eta>\eta_c=2.25$, OAM transfer to BEC can be achieved. For $\eta<\eta_c$, BEC cannot handle OAM even for a small time. We note that, 3-fold ordering does not change much in the three graphics $\eta$=1.5, 1.75 and 2.0; but changes apparently at $\eta$=2.25 and 2.5. Visibility also does not change much at the three graphics for $\eta$=2.75, 3 and 3.5.}
\label{fig7}
\end{center}
\end{figure*}

In Fig.~\ref{fig6}, we observe that OAM transfer to the BEC can be achieved only after a critical pump strength $\eta>\eta_c$ depending on the actual values of the diagonal coupling ($U_0$), collision ($g_s$) and damping of the scattered field ($\kappa$) \cite{nagy2008self}. $\ell$-fold rotational ordering in the density can appear only after $\eta > \eta_c$, where OAM transfer to the BEC becomes possible~\cite{PS}. The 3-fold ordering in the BEC profile becomes visible for $\eta > \eta_c$, see Fig.~\ref{fig7}.

In the absence of damping, $\kappa=0$, we clearly observe the staircase of plateaus at $\langle\hat{L}_z\rangle=1$ and $\langle\hat{L}_z\rangle=2$ horizontal lines. This resistance against the the OAM transfer is due to the rotationless nature ---unless multiplies of $N\hbar$ OAM is gained \cite{pethick2002}--- of the BEC. When we introduce damping, these plateaus become unobservable.

In Fig.~\ref{fig6}, we observe a gradual increase in the OAM which is managed to be transferred to the BEC. Note that: when we say OAM transfer to the BEC, for the case of non-integer $\langle \hat{L}_z\rangle$ values, we mean that BEC  can have this transient OAM values in real time dynamics only for a short time unlike quantized vorticity. We underline that, for $\eta < \eta_c$, BEC do not support OAM even for a short time.

Despite the gradual increase in $\langle\hat{L}_z\rangle$, the visibility of the $\ell$-fold rotational ordering appears quite sharply just after $\eta>\eta_c \simeq$ 2.25. In Fig.~\ref{fig7}, we observe that between $\eta$=1.50-2.25, visibility of the 3-fold ordering remains almost constant, while for $\eta>\eta_c\simeq$ 2.25 3-fold ordering appears and remains almost constant between $\eta=$ 2.75-3.50. Therefore, onset of the OAM transfer to the BEC is accompanied by a 3-fold rotational symmetric ordering~\cite{PS}.

\begin{figure}
\begin{center}
\includegraphics[width=8.5cm]{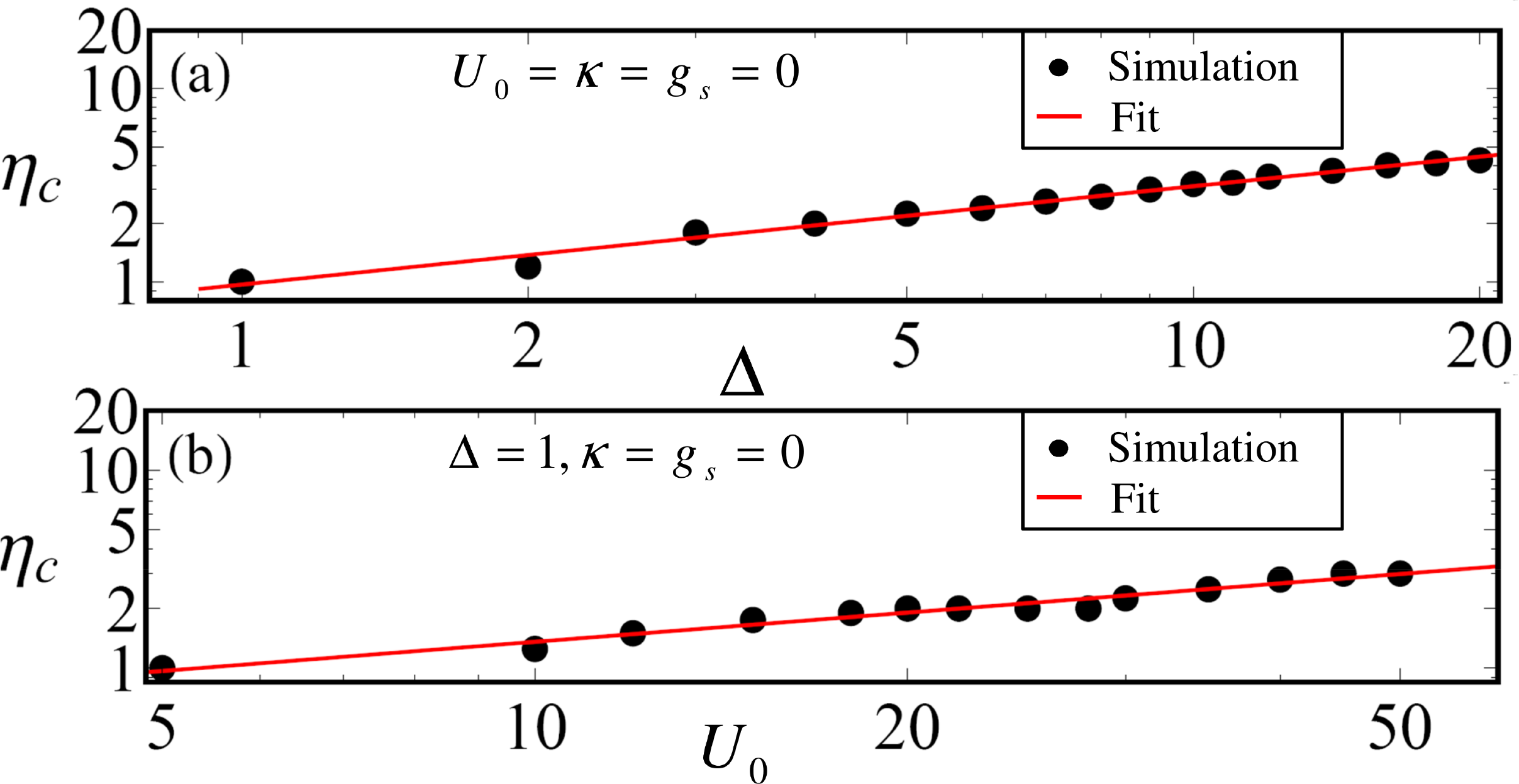}
\caption{Dependence of the crittical pump strength on the (a) end-fire mode frequency and (b) on the diagonal atom-field coupling for large values of $U_0$. The slopes of the log-log plots are 0.509 and 0.489, respectively, which indicates the behavior $\eta_c\sim \Delta^{1/2}$ and $\eta_c\sim U_0^{1/2}$ as predicted by the model of Nagy {et al.} \cite{nagy2008self,nagy2010} }
\label{fig8}
\end{center}
\end{figure}

Even though the complexity of the functions determining the superradiant emission and the recoiled BEC prevents us from establishing a simple mapping, presented in Ref.s~\cite{nagy2008self,nagy2010,esslinger2010}, in Fig.~\ref{fig8} we show that $\eta_c$ follows a form similar to the one predicted by Nagy {\it et al.}~\cite{nagy2008self}. In Fig.~\ref{fig8}, we plot $\log\Delta$-$\log\eta_c$ and $\log\Delta$-$\log\eta_c$. We observe the $\eta_c\sim \Delta^{1/2}$ for $U_0=g_s=0$ and $\eta_c\sim U_0^{1/2}$ behaviors predicted by Ref.~\cite{nagy2008self}. According to Ref.~\cite{nagy2008self}, $\eta_c\sim U_0^{1/2}$ behavior should be observed for the large values of $U_0$ since $\Delta$=1 can be neglected in this regime.

\section{SUMMARY}
In summary, we observe that BEC reorganizes to $\ell$-fold rotational symmetry above a critical pump strength $\eta_c$. For $\eta < \eta_c$, OAM transfer to the BEC does not appear. When $\eta > \eta_c$, OAM transfer~\cite{PS} and $\ell$-fold rotational symmetry show up mutually. For excess OAM transfer, $\ell$ number of single-charge vortices appear at the arms of the $\ell$-fold symmetric structures. 
    
In the time evolution, these structures appear after a sharp scattering peak, Fig. 2(b). The temporal width of this peak follows the superradiant characteristics introduced in Refs.~\cite{mandel&wolf_book,bonifacio1975cooperative}, see Fig.~\ref{fig3}.

In this paper, we do not trap the SR scattered pulse in a cavity in the x-y directions. However, if such a cavity would be present linear ordering \cite{esslinger2010} and rotatory ordering would be expected to emerge together ---possibly vortices distributed in a different pattern among the linearly ordered structure.

\section*{Acknowledgement}
The authors thank L. You and L. Deng for many useful discussions and acknowledge the financial support from TUBITAK Projects Grant No. 112T927 and 114F170. \"{O}.E.M. acknowledge support from TUBITAK Project Grant No. 112T974.

\begin{widetext}
\appendix
\section{Equations of Motion} \label{sec:EOM}
Equations of motions transforms to the following in the scaled form. 
\begin{eqnarray}
i \frac{d F(x,y)}{d t} =(-i\kappa+\Delta) F(x,y) + \eta |f(x,y)|^{2}  F_{L}(x,y) \int |Z(z)|^{2} dz + 2 U_{0} F(x,y) |f(x,y)|^{2} \int  |Z(z)|^{2} dz
\label{App_A1}
\end{eqnarray}
\begin{eqnarray}
i \frac{d f(x,y)}{d t} &=& \eta \left(F_{L}(x,y) F^{*}(x,y) + F^{*}_{L}(x,y) F(x,y)\right) f(x,y) \int |Z(z)|^{2}  dz  \nonumber 
\\
& &+ 2 U_{0} f(x,y) |F(x,y)|^{2} \int |Z(z)|^{2} dz  + \bar{g}_{s}  |f(x,y)|^{2} f(x,y) \int |Z(z)|^{4} dz \nonumber
\\
& & - \frac{1}{2} \left[ \left( \partial_x^2  + \partial_y^2 \right) f(x,y) \right] \int |Z(z)|^{2} dz
- \frac{1}{2} f(x,y) \int Z^{*}(z) \partial_z^2 Z(z)  dz  \nonumber 
\\ 
& &  + \frac{1}{2} (x^{2} + y^{2})f(x,y) \int |Z(z)|^{2} dz + \frac{1}{2} \left(\frac{\omega_{z}}{{\omega_{r}}}\right)^{2} f(x,y) \int z^{2} |Z(z)|^{2}  dz,
\end{eqnarray}
\begin{eqnarray}
i \frac{d Z(z)}{d t} &=& \eta  Z(z) \int \left( |f(x,y)|^{2} F_{L}(x,y) F^{*}(x,y) + |f(x,y)|^{2} F^{*}_{L}(x,y) F(x,y) \right) d^{2}r  \nonumber \\
 & & + 2 U_{0}  Z(z) \int |f(x,y)|^{2} |F(x,y)|^{2} d^{2}r   + \bar{g}_{s} |Z(z)|^{2} Z(z) \int |f(x,y)|^{4} d^{2}r \nonumber
\\ 
& &- \frac{1}{2} Z(z) \int  f^{*}(x,y) \left( \partial_x^2 + \partial_y^2 \right) f(x,y) d^{2}r  
- \frac{1}{2} \partial_z^2 Z(z) \int |f(x,y)|^{2} d^{2}r + \frac{1}{2} Z(z) \int (x^{2} + y^{2}) |f(x,y)|^{2} d^{2}r \nonumber 
\\ & &  + \frac{1}{2} \left(\frac{\omega_{z}}{{\omega_{r}}}\right)^{2} z^{2} Z(z)  \int |f(x,y)|^{2} d^{2}r.
\end{eqnarray}

%
\end{widetext}

\bibliography{bibliography}

\end{document}